\documentclass{article}
\title{Tachyon field in cosmology}
\author{H. K. Jassal \thanks{Email: hkj@iucaa.ernet.in} \\
{\it Inter-University Centre for Astronomy and Astrophysics}, \\
{\it Post Bag 4, Ganeshkhind, Pune-411007, India.}}

\begin{document}
\maketitle
\begin{abstract}
We study cosmological effects of homogeneous tachyon field as dark
energy.
We concentrate on two different scalar field potentials, the inverse
square potential and the exponential potential. 
These models have a unique feature that the matter density parameter
and the density parameter for tachyons remain comparable for a large
range in redshift. 
It is shown that there exists a range of parameters for which the
universe undergoes an accelerated expansion and the evolution is
consistent with structure formation requirements.
For a viable model we require fine tuning of parameters comparable to
that in $\Lambda CDM$ or in quintessence models.
For the exponential potential, the accelerated phase is followed by a
phase with  $a(t)\propto t^{2/3}$ thus eliminating a future horizon.
\end{abstract}

This report is based on recent work in collaboration with J. S. Bagla
and T. Padmanabhan \cite{bjp}.
In this paper, we construct cosmological models with homogeneous
tachyon matter \cite{asen} to provide the dark energy component which
drives acceleration of the universe (for a recent review of dark
energy models see \cite{paddy_rev}).
We  assume that the tachyon matter coexists with normal nonrelativistic
matter and radiation (for other work on cosmological aspects of
tachyon field, see\cite{cosmos}).   
For a spatially flat universe, the Friedman equations are 
\begin{eqnarray}
\label{eq:frw}
\left(\frac{\dot a}{a}\right)^2 = \frac{8 \pi G}{3} \rho,~~~~
\frac{\ddot{a}}{a}= -\frac{4 \pi G}{3} (\rho + 3p) \\ \nonumber
\end{eqnarray}
where $\rho=\rho_{\mathrm NR} + \rho_{\mathrm R} + \rho_{\phi}$, with
respective terms denoting nonrelativistic, relativistic and tachyon
matter densities.

For the tachyon field $\phi$ the energy density and pressure are given
by
\begin{eqnarray}
\label{eq:tach}
\rho_{\phi}=\frac{V(\phi)}{\sqrt{1-\dot{\phi}^2}},~~~~
p_{\phi} = -V(\phi) \sqrt{1-\dot{\phi}^2} \\ \nonumber
\end{eqnarray}
\noindent The equation of state for the tachyon  is $p=w\rho$ with
$w=\dot{\phi}^2 - 1$.
The evolution of the scalar field is described by 
\begin{equation}
\label{eq:sc_motion}
\ddot{\phi} =  - (1-\dot{\phi}^2)  \left[ 3H \dot{\phi} +
\frac{1}{V(\phi)} \frac{dV}{d \phi}\right]
\end{equation}

We  discuss cosmological models with two different potentials.   
The first one is an inverse square potential given by
$
V(\phi) =\frac{n}{4 \pi G} \left(1 - \frac{2}{3n}\right)^{1/2} \phi^{-2}.
$  
The above potential leads to the power law expansion $a(t)=t^n$ if
$\phi$ is the only source \cite{tptachyon,tptirth}. 
The term in the square bracket in the scalar field equation vanishes
in the asymptotic limit when tachyons dominate and $\phi 
\propto t$.  
All initial conditions eventually converge to this asymptotic solution. 
If normal matter or radiation dominates, $\dot\phi$ stays close to the
transition point $2/(3H\phi)$, unless we start the field very close to
$\dot\phi^2 = 1$.   
For  a viable model with matter domination at high redshifts and
an accelerating phase at low redshifts, we need to start the tachyon
field such that $\dot\phi \approx 1$ , and the field
$\phi$ is very large.     

The second form of potential is the exponential one with $V(\phi) =
V_0 e^{-\phi/\phi_0}$.    
In a radiation dominated or a matter dominated universe, $H(t)$ is a
monotonically decreasing function of time.   
As $H(t)$ keeps decreasing, $\dot\phi$ increases slowly and
asymptotically approaches unity.  
However, in a tachyon dominated scenario, the universe expands rapidly
and $H(t)$ varies much more slowly than in the matter dominated or 
in the radiation dominated era.  
Thus $\dot\phi$ changes at a slower rate but it still approaches unity
and hence we eventually get a dust like equation of state for the
tachyon field. 
Whether the evolution undergoes an accelerating phase or not depends
on the initial values of $\dot\phi$, $\phi_0$ and $\Omega_\phi$.
The present values of density parameter for non-relativistic matter and
for tachyons fix the epoch at which tachyons start to dominate the
energy density.
The parameter $\phi_0$ sets the time when $\dot\phi$ approaches unity
and the asymptotic dust like phase for tachyons is reached and
the duration of the accelerating phase is fixed by the initial value
of $\dot\phi$.
If this initial value is very close to unity, it
departs very little from it and if it starts far away from unity, the
equation of state for the tachyon field leads to a significantly long
accelerating phase.  
It is possible to fine tune the evolution  by choosing a sufficiently
large value for $\phi_0$, so that $\dot\phi$ is much
smaller than unity even at the present epoch, and by requiring that
the tachyon field starts to dominate the energy density of the
universe at the present epoch.  
For the exponential potential, the accelerating phase is a transient between the
matter dominated era and the tachyon dominated era both with $a(t) \propto
t^{2/3}$.  
These models have the attractive feature that they do not
asymptotically approach de  Sitter-like universe and hence do not
possess a future horizon (for other attempts see \cite{nohorz}).  

A comparison of these models with supernova type Ia data shows that
the models are consistent with observations for a wide range of
parameters.  
Therefore it is difficult to make a definite statement about
constraints on the parameters. 
To further check the viability and to constrain the range of
parameters, we study structure formation in tachyon models.
The unique feature in tachyon models is that at high redshift, matter
density parameter does not saturate at unity for all the models.  
This holds  true for both the potentials.
For models where the matter density parameter does not reach unity,
the growth of perturbations is slow. 
The models in which density parameter is unity at high redshifts, the
growth of perturbations is closer to that in the $\Lambda$CDM model.
The slower growth of perturbations implies that {\it rms} fluctuations
in mass distribution were  larger at the time of recombination as
compared to conventional models.  
This will have an impact on the temperature anisotropies in the
microwave background in these models.

The density parameter for matter is almost a constant at 
high redshifts ($3 < z < 10^3$), therefore in the linear limit we can
solve for the rate of growth of density contrast.
The equation for the density contrast is given by 
\begin{equation}
\ddot{\delta}+2 \frac{\dot{a}}{a} \dot{\delta}=4 \pi G \rho \delta
\end{equation}
where $\delta=(\rho - \bar{\rho})/ \bar{\rho}$, the factor
$\bar{\rho}$ being the average density.
The density contrast increases as $\delta \propto t^m$ where
$m=(1/6)(\sqrt{1 + 24   \Omega_{M}} - 1)$.   
The rate of growth slows down once the universe begins to approach an
accelerated phase from the matter dominated phase
and comes to a halt when the universe starts accelerating.  
This behavior is similar to what happens in most models with late time
acceleration of the universe.  
In the framework of tachyon cosmology, one can indeed construct viable
models in which the growth of perturbation is very similar to that in
$\Lambda$CDM models. 
These models are confined to a narrow range of
parameters. 
For parameters lying outside this range, the perturbation growth is
slower and should have higher amplitude in the past in order to
maintain a given amplitude today and consequently the models are ruled
out by cosmic microwave background observations.  

We have shown that it is possible to construct viable models where
tachyons contributes significantly to the energy density of the
universe.  
Here we have considered models in which matter, radiation and tachyons
coexist. 
It is  shown that a subset of these satisfy the constraints on
the accelerated expansion of the universe.   
However, for the accelerating phase to occur at the present epoch, it is
necessary to fine tune the initial conditions. 
The density parameter for tachyons does not becomes negligible at high
redshifts, hence the growth of perturbations in nonrelativistic
matter is slower for most models than, e.g., the  $\Lambda$CDM model.
This problem does not exist in a small subset of models.
Given that the density parameter of tachyons cannot be
ignored in the matter dominated era, it is essential to study the
fate of fluctuations in the tachyon field.

\end{document}